\documentclass[apj]{emulateapj}

\usepackage{graphicx,natbib,color}

\shorttitle{Electron acceleration and magnetic fluctuations in a solar flare loop}
\shortauthors{E.P. Kontar, I.G. Hannah and N.H. Bian}

\begin{document}


\title{Acceleration, magnetic fluctuations and cross-field transport of energetic electrons in a solar flare loop}


\author{E.P. Kontar, I.G. Hannah and N.H. Bian}
\affil{School of Physics and Astronomy,
University of Glasgow, G12 8QQ, UK}





\begin{abstract}
Plasma turbulence is thought to be associated with various
physical processes involved in solar flares, including magnetic reconnection, particle
acceleration and transport. Using Ramaty High Energy Solar Spectroscopic Imager
({\it RHESSI}) observations and the X-ray visibility analysis, we determine the spatial
and spectral distributions of energetic electrons for a flare (GOES M3.7 class,
April 14, 2002 23$:$55~UT), which was previously found to be consistent with a reconnection scenario.
It is demonstrated that because of the high density plasma in the loop, electrons have to be continuously
accelerated about the loop apex of length $\sim 2\times 10^9$cm and width $\sim 7\times 10^8$cm.
Energy dependent transport of tens of keV electrons is observed to occur
both along and across the guiding magnetic field of the loop.
We show that the cross-field transport is consistent with the presence of magnetic turbulence
in the loop, where electrons are accelerated, and estimate the magnitude of the field line diffusion
coefficient for different phases of the flare. The energy density of magnetic fluctuations
is calculated for given magnetic field correlation lengths and is larger
than the energy density of the non-thermal electrons. The level of magnetic fluctuations peaks
when the largest number of electrons is accelerated and is below detectability
or absent at the decay phase. These hard X-ray observations provide the first observational evidence
that magnetic turbulence governs the evolution of energetic electrons
in a dense flaring loop and is suggestive of their turbulent acceleration.
\end{abstract}



\section{Introduction}
In the standard flare scenario, the magnetic energy stored in the solar corona
is released accelerating particles. However, despite decades of effort, the exact
mechanisms remain elusive. It is still unclear how and where exactly
the particles are energized.
Hard X-ray (HXR) imaging spacecrafts such as Yohkoh/HXT \citep{1991SoPh..136...17K}
and RHESSI \citep{2002SoPh..210....3L} are particularly useful
for the spatial and spectral diagnostics of energetic particles \citep[][]{1999Ap&SS.264..129S,2002SSRv..101....1A,2003AdSpR..32.1001L,2005AdSpR..35.1675B,2006SSRv..124..233L,2010ApJ...712.1410T}.
\citet{1994Natur.371..495M,1995PASJ...47..677M,2004ApJ...612..546S,2008ApJ...676..704L,2009ApJ...706.1438J}
have reported observations of a non-thermal HXR source above
the top of an X-ray loop, suggesting coronal energy release via
magnetic reconnection \citep[see][as a review]{2002A&ARv..10..313P}.

Along with magnetic reconnection, another important element
of the standard flare model is magnetohydrodynamic (MHD)
turbulence as it is believed to play a key role in a number
of processes, from triggering fast reconnection to particle
acceleration and transport. A considerable number of particle acceleration
theories have been developed that require the presence of magnetic
fluctuations in the flaring loop. Assuming a high level of MHD
waves, it has been shown that stochastic acceleration
can effectively accelerate a large number
of both electrons and ions \citep{1987SoPh..113..195M,1992ApJ...398..350H,1994ApJS...90..623M,
1997JGR...10214631M,2008SSRv..134..207P,2009ApJ...692L..45B,2010A&A...519A.114B,2010ApJ...712L.131P}.
These particle acceleration models include both resonant
and non-resonant interaction between the particles and
various plasma waves. Waves and turbulence can be associated either
with current sheets during magnetic reconnection
\citep{1987ApJ...317..900C,1997ApJ...485..859S,2006A&A...452.1069L}
or with reconnection outflows \citep{1994ApJ...425..856L}.
As the magnetic field in the solar corona is weak compared
to the photosphere, direct measurement of even the mean field in the
corona is problematic. Therefore, the fluctuating components of the coronal magnetic fields
and their relation to energetic particles remain unknown. In addition,
the location and physics of turbulent acceleration within the general picture
of magnetic energy release is still awaiting an observational basis.

In this paper, using RHESSI X-ray observations and the visibility
analysis technique, we show, for the first time, the presence of energy
dependent cross-field transport of energetic electrons
in a flaring loop. We estimate the relative magnitude and energy density
of magnetic fluctuations which is required to explain these observations.
We show that these HXR observations provide a quantitative basis for the turbulence measurements
and its governing role in the evolution of energetic electrons
and are suggestive of turbulent electron acceleration in this loop.

\section{Lightcurves, spectra, and spatial distribution of X-rays from a flaring magnetic loop}

\begin{figure}\centering
 \includegraphics[width=0.9\columnwidth]{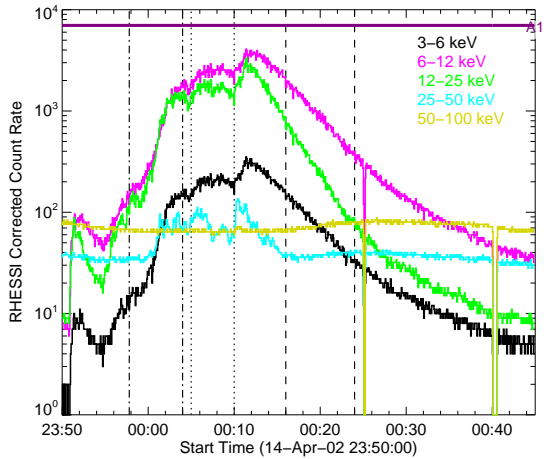}
\caption{Time profile in different RHESSI energy ranges
 of the 2002-04-14 event starting near 23:55 UT.
 The vertical lines indicates the three time
 periods for which the spectra, images and visibility forward fitting
 were performed.}
\label{fig:ltc}
\end{figure}

\begin{figure}\centering
 \includegraphics[width=0.9\columnwidth]{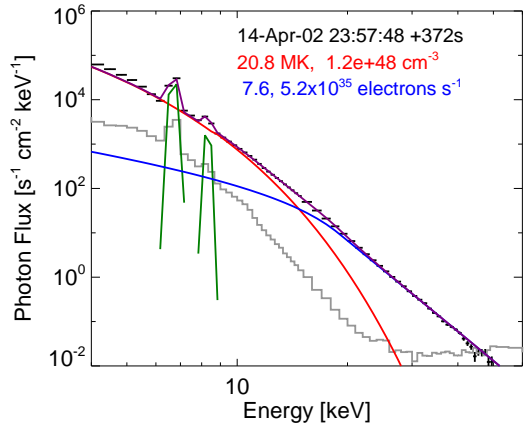}
 \includegraphics[width=0.9\columnwidth]{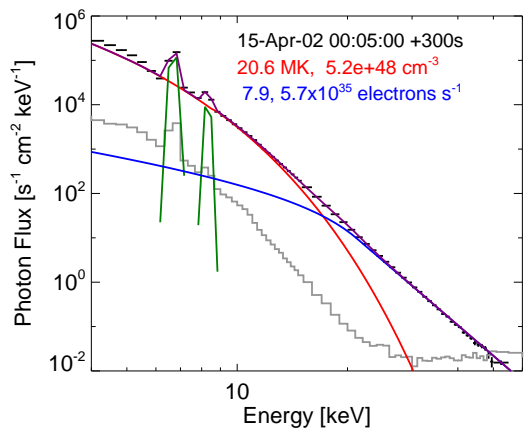}
 \includegraphics[width=0.9\columnwidth]{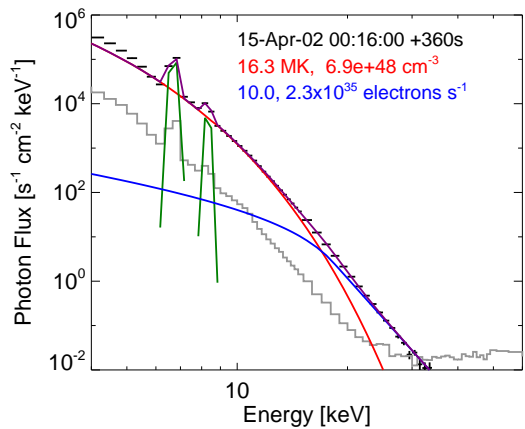}
 \caption{The spatially integrated RHESSI X-ray spectra (black data points)
 for the three time intervals during the flare.
These spectra were obtained using all front detectors excluding 2 and 7
\citep[see][for details]{2002SoPh..210...33S}
fitted (purple line) with an isothermal (red line)
plus thick-target nonthermal (blue line) model.
The pre-flare background spectra
are shown in grey.}
\label{fig:spectr}
 \end{figure}

\begin{figure*}\centering
 \includegraphics[width=2\columnwidth]{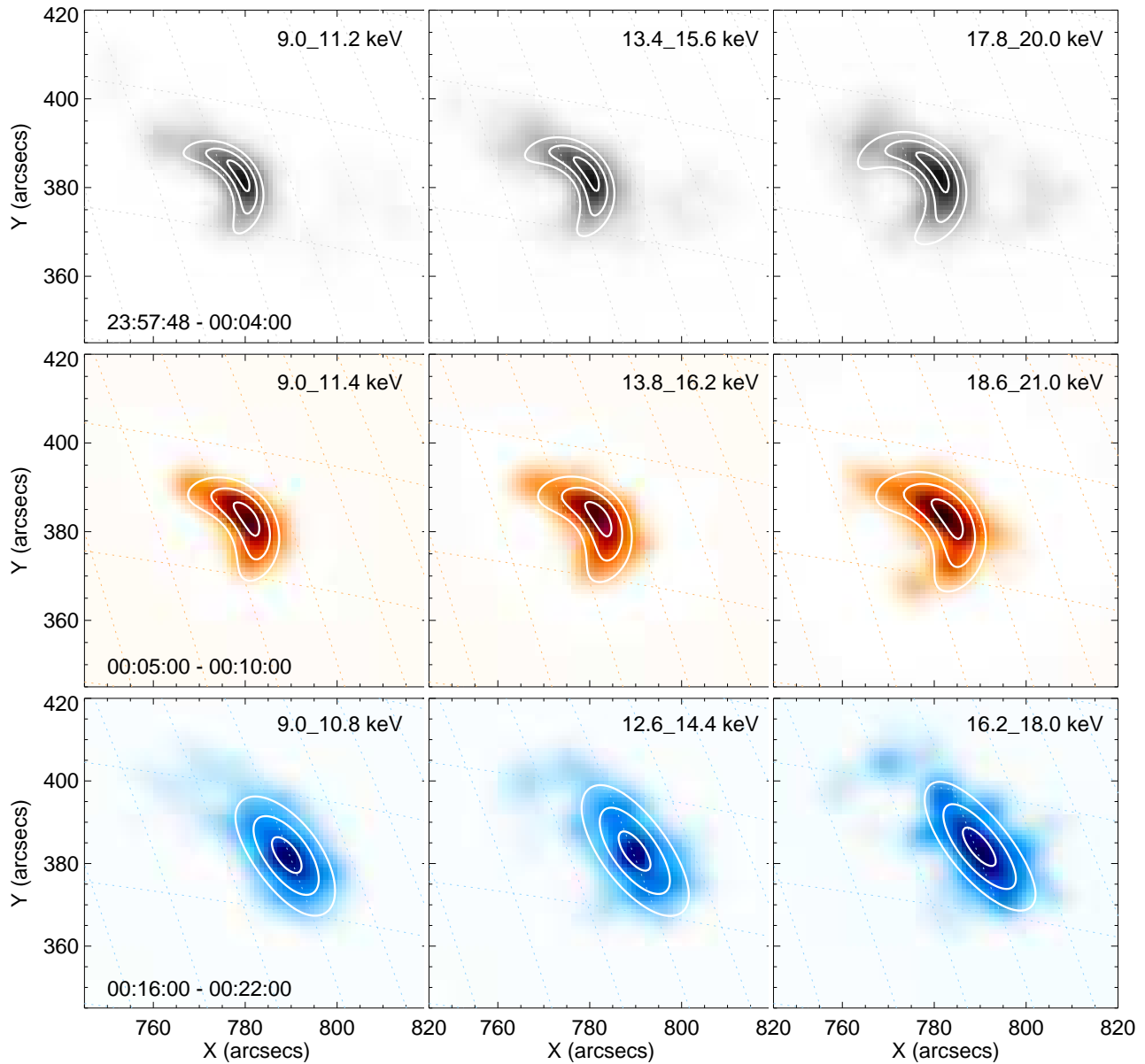}
 \caption{CLEAN image overlayed by 30, 60 and 90\% levels from the forward
fitting to X-ray visibilities. Each row indicates a different time range. Each column a different
energy range. The parameters of the loop are summarized in Table 1 and Figure \ref{fig:sizes}.}
\label{fig:images}
 \end{figure*}

We select a  GOES M$3.7$ class flare, well-observed by
RHESSI, which was previously analyzed
by \citet{2004ApJ...612..546S,2004ApJ...603L.117V,2007A&A...466..339B,2008ApJ...673..576X}.
\citet{2004ApJ...612..546S} reported the existence
of an X-ray source moving above the X-ray loop
and concluded that the observations were consistent
with the picture of the formation and development
of a current sheet between the loop-top and coronal
source.
This April 14, 2002 $23:55$~UT flare (Figures \ref{fig:ltc}-\ref{fig:images})
shows a simple loop structure
over the whole duration of the flare and, importantly, the magnetic
loop is filled with dense plasma $n\sim 10^{11}$cm$^{-3}$
\citep{2004ApJ...603L.117V}.
Unlike the majority of solar flares, which typically produce the
soft X-ray (SXR) emission in the coronal part of the loop
and bright HXR emission in the dense chromospheric
footpoints \cite[e.g.][]{2003ApJ...595L.107E,2008A&A...489L..57K},
the source of HXRs is coronal for this flare.
Because of the high plasma density in the
loop (Table \ref{table:1}), energetic electrons of
$10-20$ keV are collisionally stopped in the coronal
part of the loop and produce thick-target
emission \citep{1971SoPh...18..489B}. At the same time,
for such a dense loop, $20$ keV particles have collisional
lifetime $\lesssim0.1$s, so electrons need to be continuously
accelerated in order to explain the observed HXR
emission.

We analyse this flare in three time intervals
corresponding to the rise, peak and decay phases (see Fig \ref{fig:ltc}).
From the X-ray spectra
(Fig \ref{fig:spectr}), we determine the electron acceleration
rate  $dN/dt$ and the inverse power-law spectral
index $\delta$ of non-thermal electrons, assuming a collisional thick-target
model \citep{1971SoPh...18..489B}. We also obtain from an isothermal
component in the spectral fit \citep{2005AdSpR..35.1669H}:
the electron temperature $T_e$ and
the emission measure of thermal plasma, i.e. $EM=n^2V$,
where $n$ is the plasma
density and $V$ is the volume occupied by the thermal plasma.

During the initial phase of the flare, acceleration of electrons
up to $\sim 50$~keV is observed along with rapid heating of the plasma up to a
temperature of $20$~MK (Fig \ref{fig:spectr}).
As the flare progresses, the thermal plasma emission measure
grows while the temperature slightly increases reaching $23$~MK.
The acceleration rate $dN/dt$
is found to be around $\sim 5\times 10^{35}$~electrons~s$^{-1}$.
In the decay phase of the flare, the thermal
plasma starts to cool, the spectrum of energetic electrons
steepens and the number of non-thermal electrons
drops. The main flare parameters are summarized
in Table \ref{table:1}.
\begin{table*}
\renewcommand{\tabcolsep}{0.0cm}
\caption{HXR magnetic loop and flare parameters for the event starting April 14, 2002 23:55UT. }
\label{table:1}      
\centering           
\begin{tabular}{cccccccccc}
 Time  &$L(10 \mbox{keV})$  & $W(10 \mbox{keV})$ & $EM$  &$T_e$ & $V$ ($\pi W^2 L/4$) & $n=\sqrt{EM/V}$ & $\delta$ & $dN/dt$ &  ${\bf B}^2_{\perp}\lambda_{\parallel}/{\bf B}^2_0$\\
   & (10$^8$cm)  & (10$^8$cm)  & (10$^{48}$cm$^{-3}$) & (MK) &  (10$^{26}$cm$^3$)& (10$^{10}$cm$^{-3}$)
    & &(10$^{35}$ elec s$^{-1}$)& (10$^{7}$cm) \\
\hline \hline
 23:57:48-00:04:00&  20.3 & 4.8	&	1.2   & 21 &1.4& 9.3& 7.3& 5.2  &2.3\\
 00:05:00-00:10:00&  22.0 & 7.5 &	5.2   & 20 &3.7 &11.8&7.8 & 5.7 &2.3 \\
 00:16:00-00:24:00&  23.5 & 11.2&	6.9   & 16 & 8.9&8.8&9.8 &2.3&  $\lesssim$0.8\\
 \hline \hline
\end{tabular}
\end{table*}

X-ray images reconstructed using the CLEAN algorithm \citep{2002SoPh..210...61H}
for the three time intervals are presented in Figure \ref{fig:images}.
A clear loop structure is observed
up to $\sim 20$ keV.
Since RHESSI measures the spatial distribution of X-rays
only indirectly, via rotating modulating collimators
\citep{2002SoPh..210...61H}, X-ray images
can suffer from their reconstruction artifacts \citep{2009ApJ...706..917P,2011A&A...526A...3B}
and should not be used to determine the spatial size of the source.
Instead, a more direct and hence reliable approach is to `stack'
the rotationally-modulated lightcurves over a number of spacecraft spin periods
to produce X-ray visibilities \citep{2002SoPh..210...61H,2007SoPh..240..241S}.
The latter are instrument-independent
2-dimensional Fourier components of the HXR source \citep{2007SoPh..240..241S}.
For a simple source geometry, as displayed in this flare,
the position and sizes can be determined unambiguously
at various energies. An additional advantage of the X-ray visibility
fitting approach is that the statistical
errors can easily be propagated to find the uncertainties associated with the inferred
parameters of the loop \citep{2008A&A...489L..57K,2009ApJ...706..917P}.   \citet{2008A&A...489L..57K,2010ApJ...717..250K}
have successfully proved the robustness and sub-arcsecond
precision of this method for HXR footpoints.
X-ray visibilities are fitted using a curved elliptical gaussian model of the flare \citep{2008ApJ...677..704H,2008ApJ...673..576X}.
This is done for a number of energy ranges between $9$-$20$~keV and for the three time
intervals (see Fig. \ref{fig:images}).
The Full Width Half Maximum (FWHM) source sizes, specifically the length $L(\epsilon)$, and width $W(\epsilon)$ and
their corresponding statistical uncertainties, are obtained and presented in Figure \ref{fig:sizes}.
Using $L$ and $W$ at energy $10$~keV, we find the volume $V=\pi W^2L/4$
occupied by the thermal plasma. Assuming a volume filling factor of unity, the plasma density
is calculated: $n=\sqrt{EM/V}$ (Table \ref{table:1}). It is found to be similar to the previous
density estimates \citep{2004ApJ...603L.117V,2007A&A...466..339B,2008ApJ...673..576X}.

As anticipated from the thick-target scenario
in a weakly changing plasma density \citep[see][for details]{2002SoPh..210..373B},
the source length, $L(\epsilon)$, in the direction parallel to the guiding
magnetic field, is a growing function of energy.
The reason is that higher energy electrons can propagate further away from the region
were they are accelerated, hence making the HXR source
larger along the guiding field. In other words, in a dense coronal source,
electrons travel a distance determined by their collisional losses.
The stopping distance of an electron emitting X-rays
of energy $\epsilon$ while moving along the guiding
field of the loop
is $r_\parallel\simeq \epsilon^2/(2Kn)$,
where $K=2\pi e^4 \ln \Lambda$, $e$ is the electron
charge, $\ln \Lambda \simeq 20$ is Coulomb
logarithm \citep{2002SoPh..210..373B}. For an energy independent
length $L_{0}$ of the acceleration region, the FWHM length of the
source is given by \citep[][]{2008ApJ...673..576X}:
\begin{equation}
L(\epsilon)=L_{0}+\alpha_{\parallel} \epsilon^{2}.
\label{eq:L}
\end{equation}
By fitting Equation ($\ref{eq:L}$) to the measured $L(\epsilon)$,
it is possible to determine $L_0$ and
$\alpha_{\parallel}$ [arcsec keV$^{-2}$]. Figure \ref{fig:sizes} shows that the length
of the acceleration region $L_0$ is around $\sim 2\times 10^9$~cm at the peak of the
flare. Moreover, since $n=(2K\alpha _{\parallel})^{-1}$, $\alpha_{\parallel}$ can be used
as an independent measure to determine the plasma density. This approach
gives plasma densities of $9.8\times 10^{10}$cm$^{-3}$, $2.9\times 10^{11}$ cm$^{-3}$,
$1.7\times 10^{11}$cm$^{-3}$ for the three time intervals
analyzed in this paper. These densities are within a factor
of two from the ones which were inferred above by using the thermal emission of
the flare (see Table \ref{table:1}). Larger densities at the
peak and decay phases of the flare can be attributed to the presence of a colder
plasma invisible in Soft X-rays (SXR) with RHESSI, as suggested in previous studies of thermal
sources \citep[e.g.][]{1998A&A...334.1112J}.

\section{Energy-dependent widths of the loop}
\begin{figure}\centering
 \includegraphics[width=9cm]{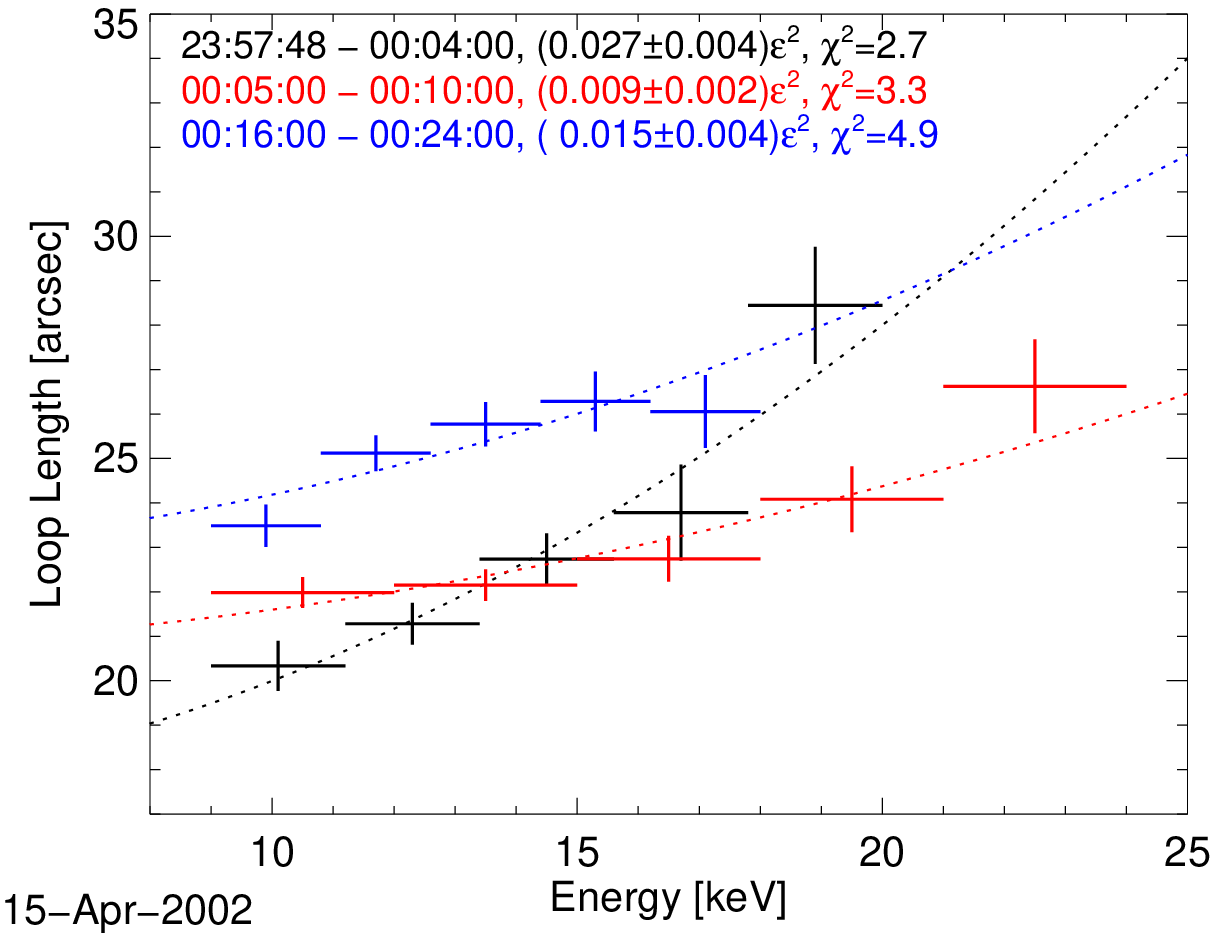}
 \includegraphics[width=9cm]{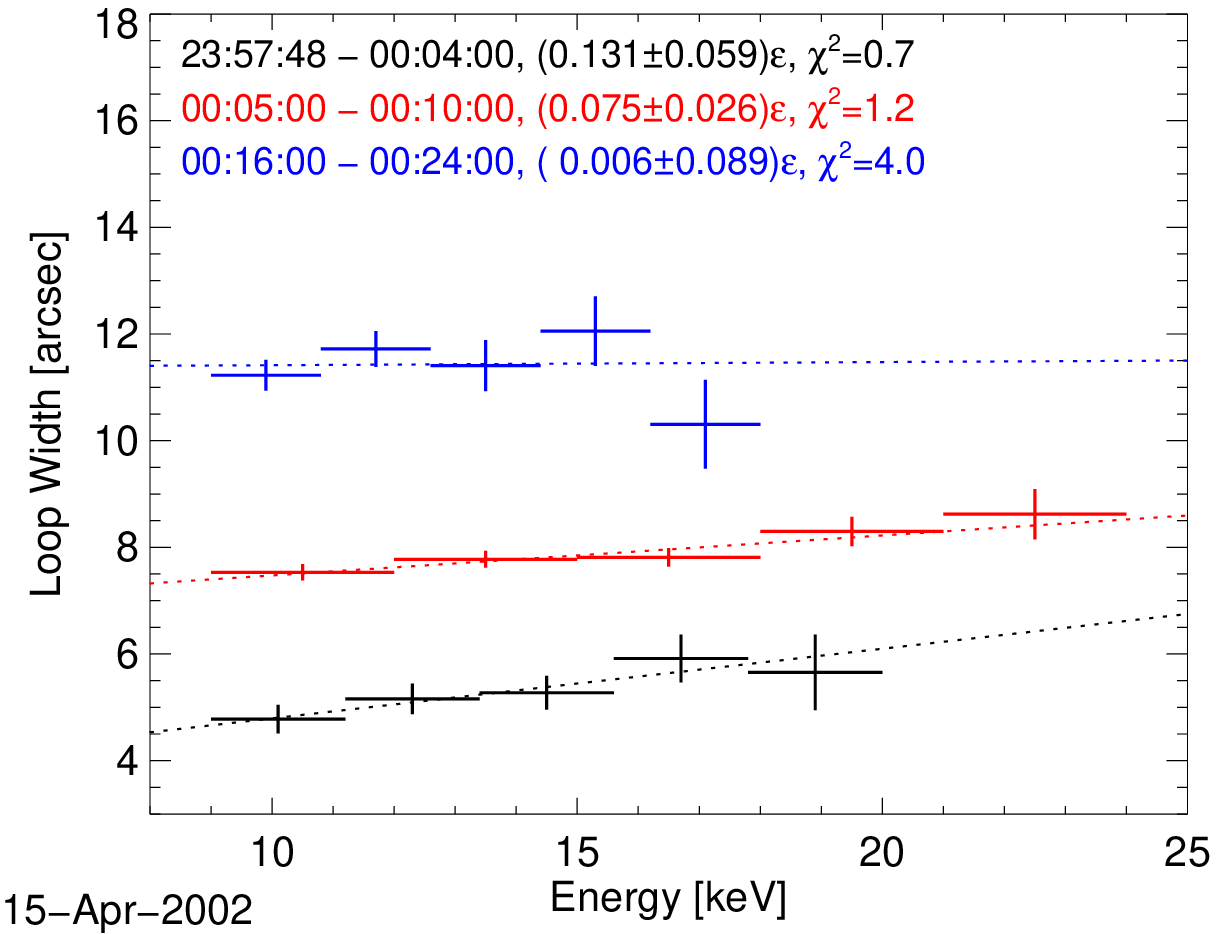}
 \caption{The FWHM loop lengths (top) and widths (bottom) found from forward fitting the
loop shape to the visibilities. The fitted lines are of the form $L(\epsilon)=L_0+\alpha _{\parallel}
\epsilon^2$ and $W(\epsilon)=W_0+\alpha_{\perp}\epsilon$ for the lengths and widths
respectively. The full $\chi^2$ quoted are calculated using the statistical uncertainties
from X-ray visibility fits. $\alpha_{\perp}$  and $\alpha _{\parallel}$ are measured
in units of [arcsec keV$^{-1}$] and [arcsec keV$^{-2}$] respectively.}
\label{fig:sizes}
 \end{figure}

The CLEAN images (Figs. \ref{fig:images}) show that the loop
width (size perpendicular to the guiding field)
is like the length, growing with energy. This is seen
quantitatively with the visibility forward fitted
as a function of energy in Figure \ref{fig:sizes}.
A growing width $W(\epsilon)$
with energy is clearly present in the first and second time
intervals (Figure \ref{fig:sizes}) but could be absent in the third.
This is particularly surprising as the guiding magnetic field
is strong enough to ensure pure field-aligned transport of energetic electrons.
Indeed, for a mean magnetic field of $\sim 150$ Gauss as
measured in this loop using radio data \citep{2007A&A...466..339B},
the Larmor radius of tens of keV electrons is $\sim2$~cm and,
therefore, the trajectories of their guiding centers coincide with the magnetic field lines.
Collisions are unimportant in producing perpendicular transport for the parameters
of the plasma, only resulting in a radial excursion (classical collisional diffusion)
of the order of the electron gyroradius in a collisional time.

However, it is well-known that particle transport across a mean magnetic field can be dramatically
enhanced, well above the collisional value, in the presence of turbulence.
The reason is the braiding of the field lines
themselves due to perpendicular magnetic fluctuations. In the presence of magnetic
fluctuations $\mathbf{B}_{\perp}$
perpendicular to the mean field $(\mathbf{B}_{0}=B_{0}\mathbf{z})$, perpendicular transport of field lines occurs which
is governed by the following equation: ${d\mathbf{r_{\perp}}(z)}/{dz}={\mathbf{B_{\perp}}(\mathbf{r_{\perp}}(z),z)}/{B_{0}}$ \citep{1978PhRvL..40...38R}.
Field lines braiding is believed to be
a dominant process governing the transport of particles
in the solar wind \citep{1966ApJ...146..480J}, in
laboratory plasmas \citep{1997PPCF...39..339B}, and in thermal coronal
loops \citep{2006ApJ...646..615G}.
It is safe to neglect the contribution of perpendicular electric fluctuations and subsequent $\mathbf{E}\times \mathbf{B}$ drift
to cross-field transport. For MHD fluctuations, the electric
field is $E_{\perp}\sim v_{A}B_{\perp}/c$ such that
the magnetic drift $\mathbf{v}_{B}=v_{\parallel}(\mathbf{B}_{\perp}/B_{0})$
dominates over the electric drift $\mathbf{v}_{E}=c\mathbf{E}_{\perp}\times \mathbf{B}_{0}/B_{0}^2$
($c$ is the speed of light and $v_{A}$ is the Alfven velocity).
Since ${v}_{B}/{v}_{E}\sim v_{\parallel}/v_{A}$, the contribution of the $\mathbf{E}\times \mathbf{B}$ drift to perpendicular
transport is negligible
for non-thermal electrons with energies above
$\gtrsim 10$~keV ($v_{\parallel}\simeq 6\times 10^9$~cm~s$^{-1}$), i.e. $v_{\parallel}\gg v_{A}$,
with $v_{A}\simeq 10^8$~cm~s$^{-1}$ in the magnetic loop.
The magnetic field line diffusion coefficient is given by the Taylor formula:
$D_{M}=(1/B_{0}^{2})\int_{0}^{\infty} d\zeta <\mathbf{B_{\perp}}(\mathbf{r_{\perp}}(\zeta),\zeta)\mathbf{B_{\perp}}(\mathbf{r_{\perp}}(0),0)>$.
Here, $<\mathbf{B}_{\perp}(\mathbf{r}_{\perp},z)\mathbf{B}_{\perp}(0,0)>$ is the two-points correlation
function and $<>$ denotes an ensemble average. In the quasi-linear approximation, the perpendicular transport
of field lines is characterized by a diffusion coefficient
$D_{M}\simeq ({B_{\perp}^{2}}/{B_{0}^{2}})\lambda_{\parallel}$,
where $\lambda_{\parallel}$ is the parallel correlation length of the perturbations
$\mathbf{B}_{\perp}$ \citep{1969ApJ...155..777J,1978PhRvL..40...38R}.
An electron moves a distance $r_{\parallel}\simeq \epsilon ^2/2Kn$ along the mean field in a collisional life-time.
Therefore, this electron also makes a radial excursion given by
$r_{\perp}=\sqrt{2 D_{M} r_{\parallel}}$, as a consequence of the radial transport of field lines.
This means that the FWHM width of the source $W(\epsilon)$ should also grow with energy.
It is the sum of the acceleration region width $W_0$ and the diffusion part due to perpendicular transport:
\begin{equation}
W(\epsilon)=W_{0}+\alpha_{\perp}\epsilon,
\label{eq:W}
\end{equation}
where $\alpha_{\perp}=\sqrt {2D_{M}\alpha _{\parallel}}$ is measured in arcsec keV$^{-1}$.
The unique feature of this expression is that it depends on the level
of magnetic fluctuations via the magnetic field line diffusion coefficient $D_{M}$.
Therefore, by measuring the energy dependent width of the flaring magnetic
loop (Figure 4), we find automatically the diffusion coefficient for the magnetic field
lines, i.e. $D_{M}=\lambda_{\parallel}{B_{\perp}^{2}}/{B_{0}^{2}}=\alpha_{\perp}^2/(2\alpha _{\parallel})$
(Table 1). As anticipated, $W(\epsilon)$ can be well fitted
with a linear fit (Figure \ref{fig:sizes})
to find $\alpha_{\perp}$ and the magnitude of magnetic fluctuations.
The magnetic diffusion coefficient $\lambda_{\parallel}{B_{\perp}^{2}}/{B_{0}^{2}}$
is changing with time and is strongest
when the acceleration rate $dN/dt$ of non-thermal electrons is at the maximum
and the spectral index of the accelerated electrons is smaller
(hardest HXR spectrum). The magnitude of fluctuations at the decay phase is
weaker, so the data cannot provide a reliable estimate,
but only an upper limit.

The energy density  of magnetic fluctuations in the loop
is $B_{\perp}^2/8\pi \simeq (B^2_0/8\pi)D_M/\lambda _{\parallel}$.
Taking the length of the acceleration region $2\times 10^9$~cm
as the upper limit for correlation length $\lambda _{\parallel}$,
the energy density of magnetic fluctuations is $\simeq 10$~erg~cm$^{-3}$
and the relative level of fluctuations $B_{\perp}/B_0\sim 0.1$.
For $\lambda _{\parallel}=2.3\times 10^7$~cm, the magnetic perturbations
follow to be strong $B _{\perp}/B_0 = 1$, and the energy density of
turbulence $\simeq 900$~erg~cm$^{-3}$. This value is close
to the energy density of the flaring plasma $3nk_BT_e \simeq 830$~erg~cm$^{-3}$
and much larger than the energy density of the non-thermal electrons
\begin{equation}
U_{nth}=\frac{1}{S}\int _{E_c}^{\infty}E\frac{F(E)}{\sqrt{2E/m_e}}\mbox{d}E,
\label{eq:E_nth}
\end{equation}
where $S=\pi{W_0^2}/4$ is the cross-sectional area of the loop,
$F(E>E_c)\propto E^{-\delta}$ is the thick-target
acceleration rate differential in energy
[electrons s$^{-1}$ keV $^{-1}$], and $E$ and $m_e$ are
the electron energy and mass respectively. Using the values
from thick-target fit and loop width measurements, the
energy density of non-thermal electrons, $U_{nth}$ is about $11$, $6$, and
$1$~erg~cm$^{-3}$ for the three time intervals respectively.
Therefore, we can conclude that the energy density of the magnetic turbulence is
sufficient to energize at least non-thermal electrons and possibly
the whole flare.

\section{Discussion and conclusions}

X-ray visibility based analysis has allowed us to infer the energy
dependent source size for a well observed flaring loop. Both
the length and the width of the loop are found to be larger
at higher energies. This is in clear contrast with HXR footpoints,
which show the opposite trend; higher energy sources appear smaller than lower energy sources \citep{2008A&A...489L..57K,2009ApJ...706..917P,2010ApJ...717..250K}.
An energy dependent growth of the loop length is expected from the collisional precipitation
of energetic electrons along field lines \citep{2002SoPh..210..373B,2008ApJ...673..576X}.
The high density of the loop implies that the electrons
are accelerated inside the loop region
with the characteristic size $2\times 10^9$~cm.
At the same time, the size of the loop perpendicular to the strong magnetic
field is an increasing function of energy. The increase of the loop width implies strong
cross-field transport
of non-thermal electrons and is found to be consistent with the diffusion of magnetic
field lines. This allows us to determine the level and locations
of magnetic fluctuations in the flaring loop.
For comparison, we note that the magnetic diffusion coefficient
$D_M=\lambda_{\parallel}{B_{\perp}^{2}}/{B_{0}^{2}}$
measured {\it in situ} in the solar wind, where $B _{\perp}/B_0 \sim 1$
and $\lambda _{\parallel}\sim 10^{11}$~cm \citep{1986JGR....91...59M}
is about four orders of magnitude larger than our values presented
in Table \ref{table:1}.

We note that although the length and width of the flare
are likely to be governed by parallel transport and diffusion of the magnetic
field lines, equations (\ref{eq:L},\ref{eq:W}) for the width
and the length are approximations. The simultaneous treatment of acceleration,
energy losses and cross-field transport is needed to assess the particle dynamics more
accurately. This might lead to super or sub-diffusive cross-field
transport of electrons and cannot be ruled out by the current data,
so the radial excursions can scale as $r_{\perp}\propto \epsilon^{\beta}$,
with $\beta$ close to one. Nevertheless, the width-energy measurements
provide a powerful and unique tool to measure magnetic fluctuations
in a flaring plasma. Importantly, the energy density of magnetic
fluctuations appears to be energetically significant and is found to be
larger than the energy density of non-thermal particles.

The temporal and spatial analysis by \citet{2004ApJ...612..546S}
of the weaker source above the dense loop analyzed here
suggests a picture consistent with formation of a reconnecting
current sheet between the loop top and the coronal source.
Assuming this reconnection scenario, our observations
suggest the presence of MHD turbulence in the loop under
the reconnection site. During the event, the length of acceleration site
increases only by $\sim 10$\%, while the width of the bright HXR
loop $W_0$ grows with time more than a factor of $\sim 2$,
which could be attributed to continuous piling up of newly
reconnected lines. The plasma mode responsible for the perturbations
remain unknown, but these observations indicate the presence
of noticeable magnetic fluctuations during the solar flare inside the loop,
where the acceleration of electrons is happening. It is possible to speculate
that the magnetic field fluctuations are produced by magnetic
reconnection above the loop and accelerate electrons
inside the flaring loop.

\acknowledgments
This work is partially supported by a STFC rolling grant.
Financial support by the Leverhulme Trust (EPK),
and by the European Commission through the SOLAIRE
(MTRN-CT-2006-035484) and HESPE (FP7-SPACE-2010-263086)
Networks is gratefully acknowledged. The overall effort has greatly
benefited from support by a grant from the International Space Science
Institute (ISSI) in Bern, Switzerland.



\bibliographystyle{apj} 

\end{document}